# Exploring Chalcogen Influence on $Sc_2BeX_4$ (X = S, Se) for Green Energy Applications Using DFT

Ahmad Ali *[1,2], Haris Haider [1,2], Sikander Azam [3], Muhammad Talha [4], Muhammad Jawad **[3], Imran Shakir [5]

[1] Department of Physics, Abdul Wali Khan University, Mardan, 23200, Pakistan

[2] Department of Physics, Government Degree College Lahor, Swabi, Pakistan

[3] Faculty of Engineering and Applied Sciences, Department of Physics, Riphah International University, Islamabad, Pakistan

[4] Department of Physics, University of Swabi, Swabi, 23562, Pakistan

[5] Department of Physics, Faculty of Science, Islamic University of Madinah, Madinah 42351, Saudi Arabia

## Abstract

Using density functional theory, the structural, electronic, optical, and thermoelectric properties of $Sc_2BeX_4$ (X = S, Se) chalcogenides were investigated for energy applications. Both compounds are dynamically and thermodynamically stable with negative formation energies (−2.6 eV for $Sc_2BeS_4$ and −2.2 eV for $Sc_2BeSe_4$). They exhibit direct band gaps: 1.8 eV (S) and 1.2 eV (Se) via TB-mBJ, suggesting strong visible light absorption. Optical parameters reveal high static dielectric constants (9 and 16.5), peak absorption at ~13.5 eV, and reflectivity under 30%. Thermoelectric analysis shows p-type behavior with Seebeck coefficients up to $2.5 \times 10^{-4}$ V/K and electrical conductivities of $2.45 \times 10^{18}$ and $1.91 \times 10^{18}$ $(\Omega\, m\, s)^{-1}$ at 300 K. The power factors reach $1.25 \times 10^{11}$ W/K² m s, and ZT values attain 0.80 at 800 K. Debye temperatures (420 K for S, 360 K for Se) suggest low lattice thermal conductivity. These results designate $Sc_2BeX_4$ as promising candidates for photovoltaic and thermoelectric applications.

*Corresponding Author:([*aaphy12@gmail.com](mailto:aaphy12@gmail.com))([**mjawad1991@gmail.com](mailto:mjawad1991@gmail.com))

# 1. Introduction

Ternary chalcogenide compounds have emerged as a vibrant area of materials research because of their rich chemistry, tunable physical properties, and technological promise in energy conversion and optoelectronic devices. These materials, which are generally composed of transition or main-group metals combined with chalcogen elements (S, Se, or Te), combine earth-abundant and non-toxic constituents with versatile crystal structures. As a result, they have attracted widespread attention across disciplines including physics, materials science, and chemistry [1–5]. The ability of these compounds to accommodate a wide range of elemental substitutions allows researchers to engineer their electronic band structures, dielectric responses, and thermal transport behavior, which are crucial parameters in applications such as photovoltaics, photodetectors, thermoelectrics, and photocatalysis [6–8]. Their structural flexibility and the possibility of controlling band gaps through composition make them particularly relevant for modern green-energy technologies that require cost-effective, non-toxic, and stable materials. Within the broader family of chalcogenides, ternary transition-metal chalcogenides stand out because they often exhibit semiconducting behavior with band gaps that lie in the visible to near-infrared region, which is highly desirable for harvesting solar energy and for designing optical coatings or sensors. For example, Lei et al. [9] highlighted that antimony-based chalcogenides are highly anisotropic and display a range of functionalities, from photocatalysis to thermoelectric energy conversion and Li-ion batteries. Similarly, research on copper, indium, and gallium selenides has demonstrated how elemental tuning can lead to thin-film solar cell materials with improved performance and cost efficiency. Compounds such as $CuGaSe_2$ and $CuInSe_2$ are now recognized as leading thin-film absorbers because of their direct band gaps and high absorption coefficients [10].

In parallel with experimental advances, density functional theory (DFT) has become an indispensable tool for predicting and understanding the properties of unexplored chalcogenides before costly synthesis efforts are undertaken. Several recent studies have demonstrated the effectiveness of first-principles calculations in identifying promising candidates. For example, G-Said et al. [11] investigated the ternary selenide $Tl_3AsX_3$ (X = S, Se) using the TB-mBJ potential and revealed favorable optoelectronic properties such as high optical conductivity and positive refractive indices, pointing to multi-functional applications. Aparna et al. [12] applied FP-LAPW calculations to spinel $HgLa_2X_4$ (X = S, Se) and found promising thermoelectric performance, with band gaps computed using TB-mBJ in close agreement with experimental trends. Banat et al. [13] reported negative formation energies and good thermoelectric behavior for $Ba_2GeS_4$ and $Ba_2GeSe_4$, highlighting the potential of barium-based chalcogenides in energy harvesting. Qasim et al. [14] investigated $ZnCrX_2$ (X = S, Se, Te) and showed that these materials possess indirect band gaps with semiconducting nature and tunable optical characteristics. Similar computational investigations by Yaser et al. [15] on $Y_2ZnX_4$ (Y = In, Ga; X = S, Se) and by Jawad et al. [16] on $Sr_2GeX_4$ (X = S, Se) further demonstrate how systematic substitution of chalcogens can strongly influence electronic structures and thermoelectric parameters. Despite these substantial efforts, many chalcogenides remain unexplored, leaving a wide space for discovering new compounds with superior or complementary properties. Among these, the Sc$_2$BeX$_4$ (X = S, Se) system is particularly intriguing. Incorporating lightweight cations such as Scandium (Sc) and Beryllium (Be) is expected to lower mass density and influence vibrational properties, potentially enhancing phonon-related behavior such as thermal conductivity. Moreover, both Sc and Be are non-toxic elements, making these compounds attractive from an environmental and sustainability perspective. A systematic evaluation of Sc$_2$BeS$_4$ and Sc$_2$BeSe$_4$ has not yet been reported, and their structural, optical, electronic, and thermoelectric properties remain largely unknown. This lack of prior investigation underscores the novelty and scientific significance of the present work.

An important aspect that makes Sc$_2$BeX$_4$ noteworthy is the anticipated role of the chalcogen atom in tuning its properties. Replacing sulfur with selenium generally increases the polarizability of the lattice and modifies the band gap, dielectric constant, and optical spectra, which can lead to enhanced absorption in the visible region and potentially higher carrier mobility. Such trends have been observed in related systems and are particularly relevant when designing materials for green-energy applications. Understanding these chalcogen-driven modifications at the electronic and phononic levels is essential to assess the suitability of these compounds for thermoelectric and optoelectronic devices.

In light of these motivations, the present work employs a comprehensive first-principles approach to investigate Sc$_2$BeX$_4$ (X = S, Se). Using the full-potential linearized augmented plane-wave (FP-LAPW) method implemented in WIEN2k, structural optimizations are performed within the generalized gradient approximation (PBE-GGA), while band gaps and optical spectra are refined using the Tran–Blaha modified Becke–Johnson (TB-mBJ) potential[17, 18]. The dynamical stability of these compounds is examined via phonon dispersion calculations, and their thermoelectric transport properties are evaluated using the Boltzmann transport equation as implemented in the BoltzTraP code. By systematically comparing the S and Se variants, we highlight how chalcogen substitution affects band structure, phonon stability, optical coefficients, and thermoelectric performance. To the best of our knowledge, this is the first detailed theoretical investigation of Sc$_2$BeX$_4$, and the insights presented here contribute to identifying new, environmentally friendly materials for future green-energy applications.

2. **Computational Details**

The FP-LAPW method has been utilized to perform the calculations within the context of DFT [19, 20], implemented in the simulation code-named Wien2k [21]. This method has proven to be one of the most accurate approaches for the calculation of electronic structures [22-26]. The materials Sc$_2$BeX$_4$ (X=S & Se) are ternary chalcogenide compounds that exhibit a tetragonal structure with the space group number 122 (I-42D), and the visualized crystallographic structure has been developed using VESTA software, displayed in Fig. 1 [27]. The density of states, optical, and thermoelectric parameters are plotted using the Origin software [28], while the band structure plots are developed by XMGRACE [29]. The BoltzTrap algorithm is applied to reveal the transport or thermoelectric properties with the use of the BoltzTrap code [30]. The phonon stability and thermodynamic characteristics of the materials have been predicted using the CASTEP code [31]. The Sc atom is in a unit cell at the location (0.13, 0.25, 0.62), the Beryllium (Be) atom is located at (0, 0, 0), while the Sulphur (S) and Selenium (Se) atoms are located at the positions (0.06, 0.19, 0.31), respectively. The crystallographic structure of the materials is displayed in Fig. 1. The TB-mBJ and PBE-GGA approximation functions are employed separately to determine the ground state properties of the materials $Sc_2BeX_4$ (X = S, Se) [32]. GGA-PBE was used for structural optimization, while TB-mBJ was employed to improve the band-gap description. Although mBJ+U+SOC could yield even more accurate band gaps, as reported in literature [17, 18], this work focuses on trends and relative differences, and such expensive calculations are beyond the scope. The equations of TB-mBJ and PBE+GGA are shown below, respectively [33];

$$E_{xc}[\rho] = \int \rho(r)\epsilon_{xc}(\rho(r), \nabla\rho(r))\,dr \qquad (1)$$

$$V_{XC}^{mBJ} = V_{XC}^{LDA} + \nabla V_{XC}^{mBJ} \qquad (2)$$

In our calculations, the value of RMT x Kmax is considered to be 8 for the energy convergence to be reasonable, which specifies the size of the matrix for the convergence. For $Sc_2BeS_4$, the RMTs are 2.5 for Sc, 2.21 for Beryllium (Be), and 2.23 for Sulphur (S), similarly in $Sc_2BeSe_4$ the RMTs are 2.5 for Sc, 2.14 for Beryllium (Be) and 2.5 for Selenium (Se); with the atomic number of 21 for (Sc), 4 for (Be), while 16 and 34 for (S) and (Se) respectively. The following valence electrons were treated explicitly: Sc ($3d^1 4s^2$), Be ($2s^2$), S ($3s^2 3p^4$), Se ($4s^2 4p^4$). Core states were kept frozen, as implemented in the WIEN2k FP-LAPW method.

The energy and charge convergence values for SCF (self-consistent field) computations were chosen to be 0.0001 Ry and 0.001e, respectively. The values of $G_{max}$ and $l_{max}$ (angular momentum vector) in the atomic sphere were found to be 12 and 7, respectively. To ensure the excellent and accurate convergence of the total energy 1000 k-points were chosen in the irreducible first Brillouin zone. For the separation of the valence and core states, -6.0 Ry is selected as the cutoff energy. The complex dielectric function ε(ω), a frequency-dependent function described using Ehrenreich and Cohen's relations, is used to analyze the optical characteristics [19],

$$\varepsilon(\omega) = \varepsilon_1(\omega) + i\varepsilon_2(\omega) \qquad (3)$$

Due to the interaction of the fields of incident photons and electrons of materials, the optical transition (occupied and unoccupied states) takes place. The imaginary part of the dielectric function can be obtained with the help of the momentum matrix dipole, which lies between the occupied and unoccupied states. Additionally, the Kramer-Kronig relation is utilized to obtain the real component [34, 35]. The real and imaginary part of the complex dielectric function is given as;

$$\varepsilon_1(\omega) = 1 + \frac{2}{\pi} P \int_0^\infty \frac{\omega \varepsilon_2(\omega)}{\omega'^2 - \omega^2} d\omega' \tag{4}$$

Here, 'P' is the principal integral.

$$\varepsilon_2(\omega) = \frac{e^2 h'}{\pi m^2 \omega^2} \sum_{v,c} \int BZ \, |M_{c,v}(k)|^2 \delta[\omega_{cv}(k) - \omega] d^3k \tag{5}$$

$M_{c,v}(k)$ is the dipole matrix element

The optical parameters such as refractive index n(ω), the absorption coefficient I(ω), the energy loss function L(ω), and the reflectivity R(ω) are obtained from $\varepsilon_1$(ω) and $\varepsilon_2$(ω); they are shown respectively.

$$n(\omega) = \left[\frac{1}{2}\left(\sqrt{\varepsilon_1(\omega)^2 + \varepsilon_2(\omega)^2} + \varepsilon_1(\omega)\right)\right]^{\frac{1}{2}} \tag{6}$$

$$I(\omega) = \frac{\sqrt{2}\omega\left(\sqrt{\varepsilon_1(\omega)^2 + \varepsilon_2(\omega)^2} - \varepsilon_1(\omega)^2\right)^{\frac{1}{2}}}{c} \tag{7}$$

$$L(\omega) = \frac{\varepsilon_2(\omega)}{\varepsilon_1^2(\omega) + \varepsilon_2^2(\omega)} \tag{8}$$

$$R(\omega) = \left|\frac{\sqrt{\varepsilon_1(\omega) + i\varepsilon(\omega)} - 1}{\sqrt{\varepsilon_1(\omega) + i\varepsilon(\omega)} + 1}\right| \tag{9}$$

Now the various desired properties of our materials $Sc_2BeX_4$(X=S & Se) have been computed and discussed, including their band structures, density of states, and optical, thermoelectric, and thermodynamic properties in the next section below;

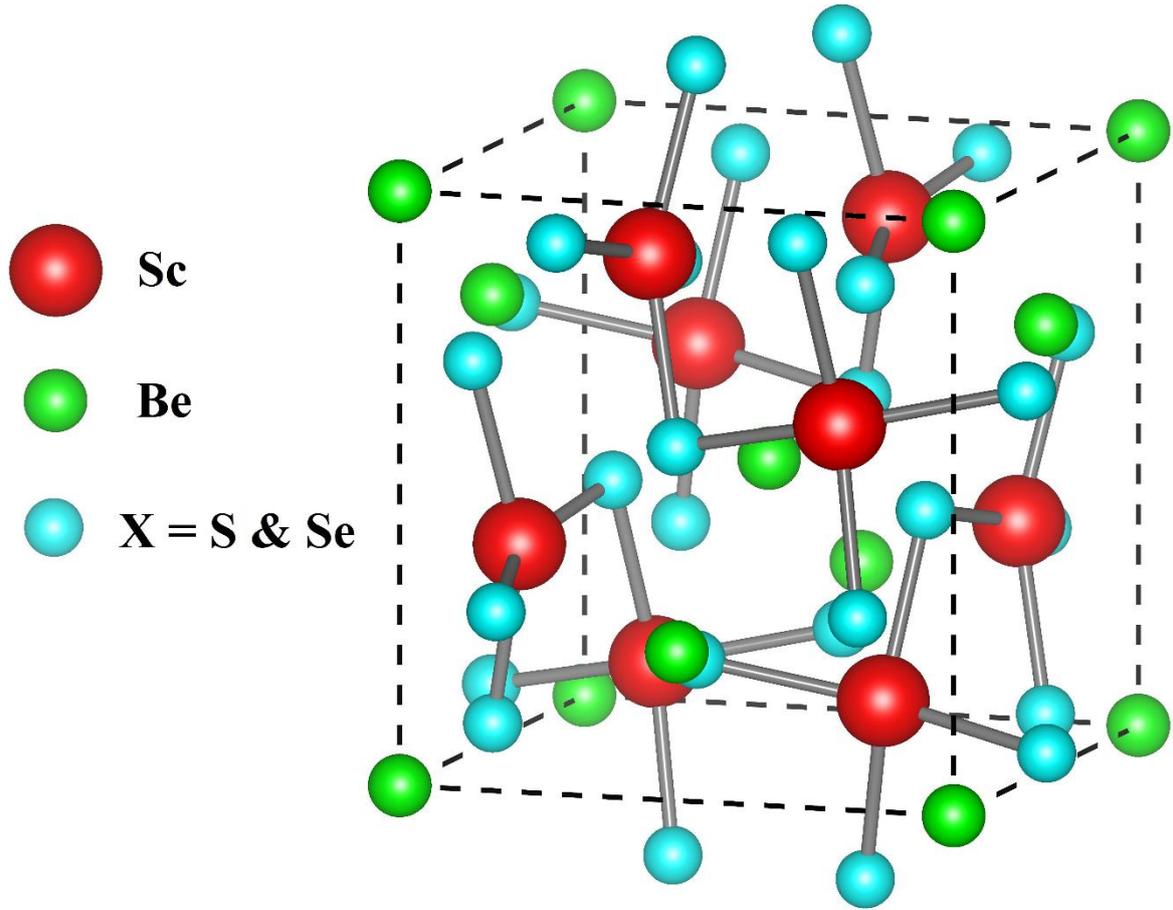

**Fig. 1.** Crystal structure of unit cells of $Sc_2BeS_4$ and $Sc_2BeSe_4$ developed by Vesta

## 3. Results and Discussions

### 3.1. Structural Stability

The structural stability of a compound can be determined by studying different stability criteria for the materials. The tolerance factor τ, defined by Goldschmidt in a modified format, is an important factor that explains the phase stability of the materials under study [36]. The modified form of the empirical relations for the tolerance factor has been defined as;

$$\tau = 0.707 x \frac{r_{Sc} + r_{S/Se}}{(r_{Be} + r_{S/Se})}$$

Where, $r_{Sc}$, $r_{Be}$, $r_S$, and $r_{Se}$ represent the ionic radii of the constituent atoms of the materials $Sc_2BeS_4$ and $Sc_2BeSe_4$ in the empirical relation. The values of the ionic radii are $r_{Sc}$ 0.745 Å, $r_{Be}$ = 0.27 Å, $r_S$ = 1.84 Å, and $r_{Se}$ = 1.98 Å [37]. The computed tolerance factors of the materials $Sc_2BeS_4$ and $Sc_2BeSe_4$, using the above empirical relation, are 0.866 and 0.856, respectively. The stability range for the cubic phase is 0.9 to 1 [38]. The values of the tolerance factor of materials reveal distortion in the cubic crystallographic phase and predict the phase shift from the cubic to the tetragonal phase of the materials, which confirms the stability of the materials in their tetragonal phase.

The formation energy analysis of the material provides details about the thermodynamic stability of the material. The formation energy of the materials can be calculated using the relation;

$$E_F = [E_{Sc_2BeX_4(X=S\ \&\ Se)} - (2xE_{Sc} + 1xE_{Be} + 4xE_{X=S\ \&\ Se})]/7$$

Here, $E_F$ is the formation energy of the materials $Sc_2BeX_4$(X=S & Se). The $E_{Sc_2BeX_4(X=S\ \&\ Se)}$, $E_{Sc}, E_{Be}$, and $E_{X=S\ \&\ Se}$ are the total energies of the materials, Sc and Be, S, and Se, respectively. These energies are obtained by carrying out the optimizations in the FM phase. The computed formation energy per atom of the materials $Sc_2BeS_4$ and $Sc_2BeS_4$ are -2.6 eV and -2.2 eV, respectively. The negative formation values of the material confirm the thermodynamic stability of the materials [39].

The dynamic stability of the materials $Sc_2BeX_4$(X=S & Se) can be understood by the phonon dispersion analysis and phonon DOS, which have been computed and displayed in Fig. 2. The range of phonon frequencies for $Sc_2BeS_4$ and $Sc_2BeSe_4$ is 0 to 15.2 THz and 0 to 13.8 THz, respectively. The phonon DOS plots also follow the phonon dispersion behavior for both materials. The positive phonon frequencies of the materials suggest that the materials are dynamically stable [40-42]. We confirm that the phonon dispersion curves were computed over the entire frequency range up to 25 THz. No imaginary modes were observed, ensuring the dynamic stability of both compounds even beyond 15 THz. This stability supports the structural robustness predicted by the tolerance factor [43-44].

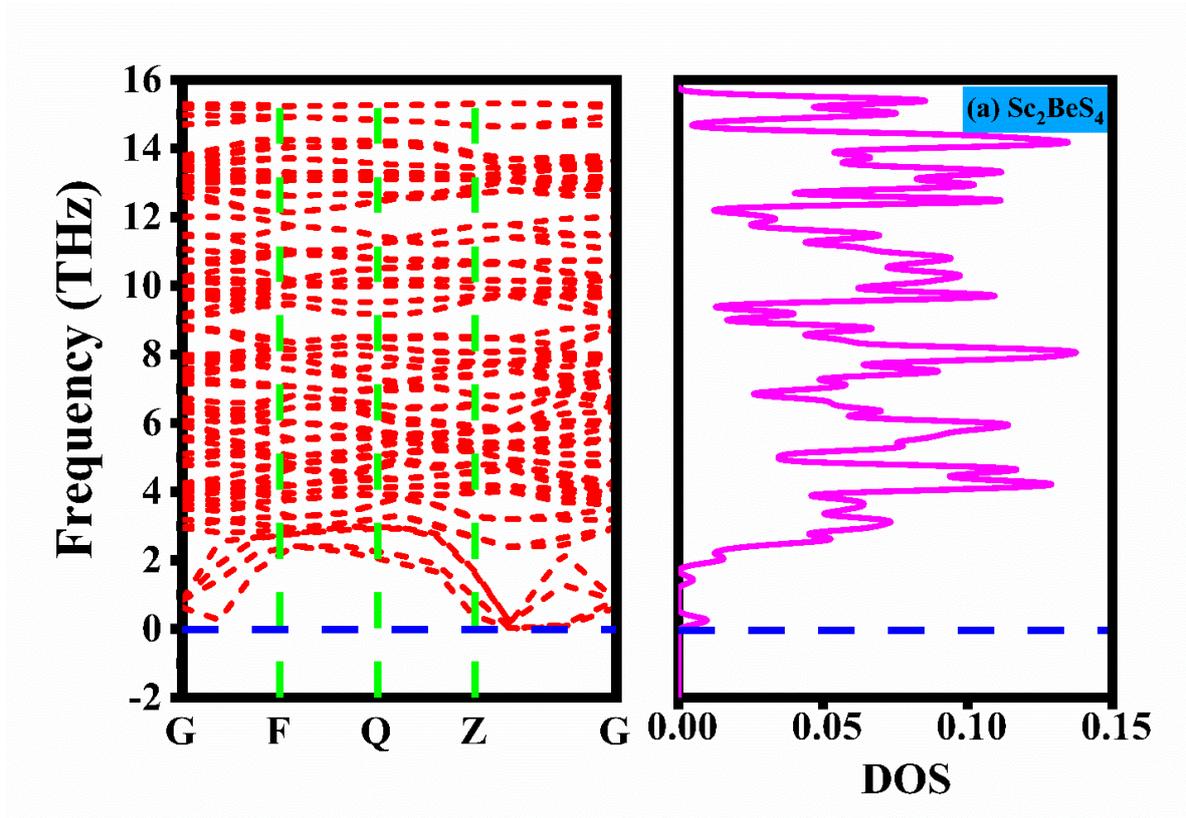

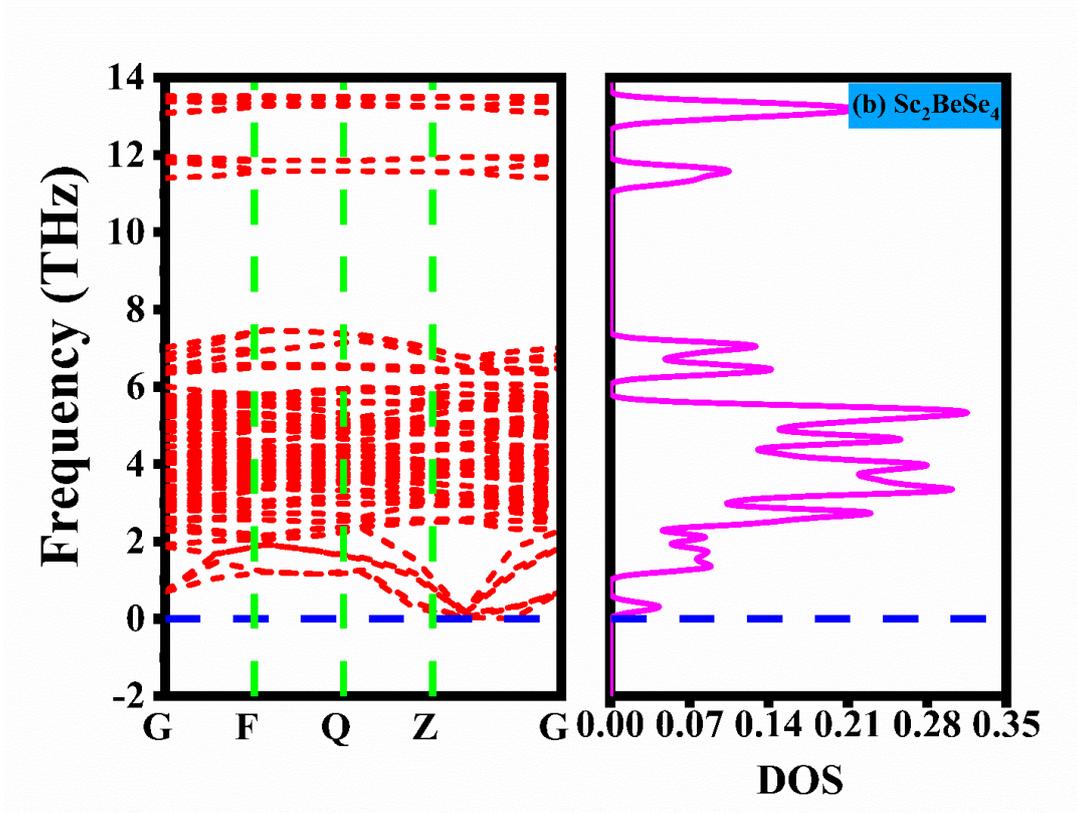

**Fig. 2:** The Phonon dispersion plots along with the DOS of (a) $Sc_2BeS_4$ and (b) $Sc_2BeSe_4$

## 3.2. The Electronic Properties

The electronic band structures together with the total and partial density of states (TDOS and PDOS) of Sc$_2$BeX$_4$ (X = S, Se) were calculated to elucidate the electronic features of these compounds. Both PBE-GGA and TB-mBJ exchange–correlation potentials were employed to provide a comparative analysis of the band gaps and the contribution of atomic orbitals. The band structures were calculated along the high-symmetry path Γ–H–N–Γ–P for both materials. The Fermi level (E$_F$) in all plots is set at 0 eV. Figure 3(a) shows that Sc$_2$BeS$_4$ exhibits a semiconducting nature with a direct band gap, where both the valence band maximum (VBM) and conduction band minimum (CBM) occur at the Γ point. The band gap is 1.4 eV using PBE-GGA and increases to 1.8 eV with TB-mBJ, which is consistent with the known tendency of TB-mBJ to provide improved band gap values for semiconductors. For Sc$_2$BeSe$_4$, Figure 3(b) shows a similar direct band gap character with slightly smaller values of 0.8 eV (PBE-GGA) and 1.2 eV (TB-mBJ). The observed reduction in the band gap when S is replaced by Se is a typical chalcogen effect, attributed to the larger polarizability and orbital extension of Se, which shifts the valence and conduction band edges and consequently narrows the band gap. These results directly demonstrate the chalcogen influence on the electronic structure: the substitution of S by Se reduces the band gap, a property that can enhance electrical conductivity and is advantageous for optoelectronic and green-energy applications. The calculated band gaps fall in the range desirable for photovoltaic absorbers, where values close to ~1.6 eV are considered optimal for single-junction solar cells [45]. The direct nature of the band gaps also indicates strong absorption in the visible region, making these materials promising for photodetectors and light-emitting devices in the UV and near-infrared ranges [46-48].

The TDOS and PDOS further clarify the orbital contributions. For Sc$_2$BeS$_4$ (Fig. 4a), the valence band region extending from 0 to –5 eV is predominantly composed of S-p states with minor contributions from Sc-d orbitals. The conduction band region beginning near 1.2 eV in PBE-GGA and 1.8 eV in TB-mBJ is primarily dominated by Sc-d and Be-s states with smaller S-p contributions. For Sc$_2$BeSe$_4$ (Fig. 4b), a similar trend is observed: the valence band from 0 to –4.8 eV is mainly contributed by Se-p states, while the conduction band from 0.8 eV (PBE-GGA) or 1.4 eV (TB-mBJ) upward is dominated by Sc-d and Be-s states with minor Se-p involvement. This orbital analysis confirms that the replacement of S by Se not only reduces the band gap but also shifts the density of states distribution near the band edges, which is consistent with observed behavior in other halide and chalcogenide perovskites. The density of states in both approximations validates the band structure results and reinforces the conclusion that both Sc$_2$BeS$_4$ and Sc$_2$BeSe$_4$ are direct band gap semiconductors with significant chalcogen-dependent electronic tuning. Such a combination of direct band gap, tunable electronic structure, and non-toxic elemental composition suggests that these materials are promising for a wide range of photonic and thermoelectric applications, including solar cells [49,50], UV–LEDs [48], and near-infrared photodetectors [47].

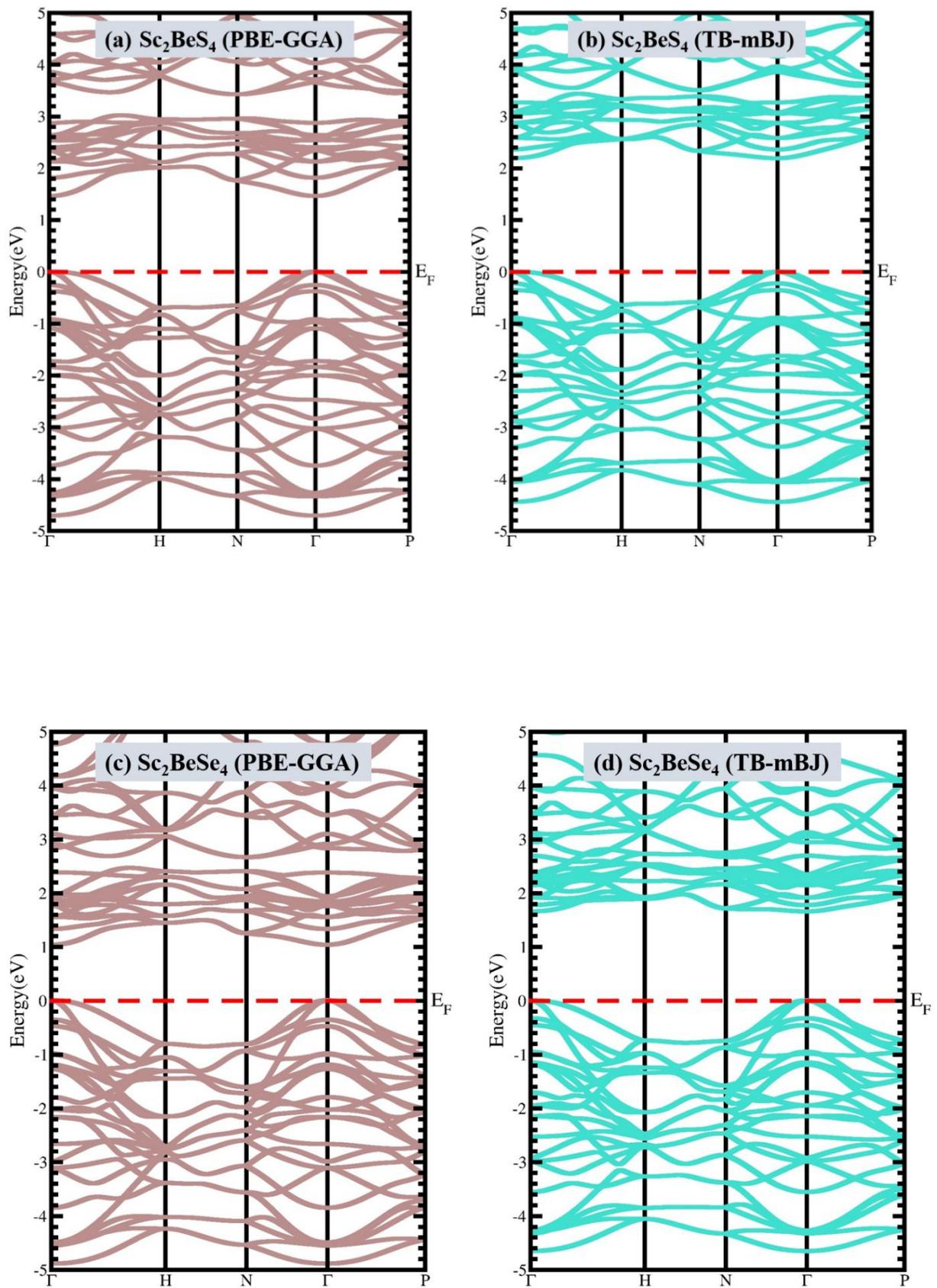

**Fig. 3.** Illustration of the electronic band structure of (a-b) Sc$_2$BeS$_4$ and (c-d) Sc$_2$BeSe$_4$ using PBE-GGA and TB-mBJ potentials.

**Table 1.** Summarize the illustration of band-gap energies of $Sc_2BeX_4$ (X = S & Se)

| Materials | Potentials | Γ-Point | H-Point | N-Point | P-Point | Difference | Nature |
|---|---|---|---|---|---|---|---|
| $Sc_2BeS_4$ | PBE-GGA | 1.4eV | 1.9eV | 2.0eV | 2.2eV | 0.4eV | Direct |
| | TB-mBJ | 1.8eV | 2.3eV | 2.4eV | 2.6eV | | |
| $Sc_2BeSe_4$ | PBE-GGA | 0.8eV | 1.5eV | 1.6eV | 2.0eV | 0.4eV | Direct |
| | TB-mBJ | 1.2eV | 1.9eV | 2.0eV | 2.4eV | | |

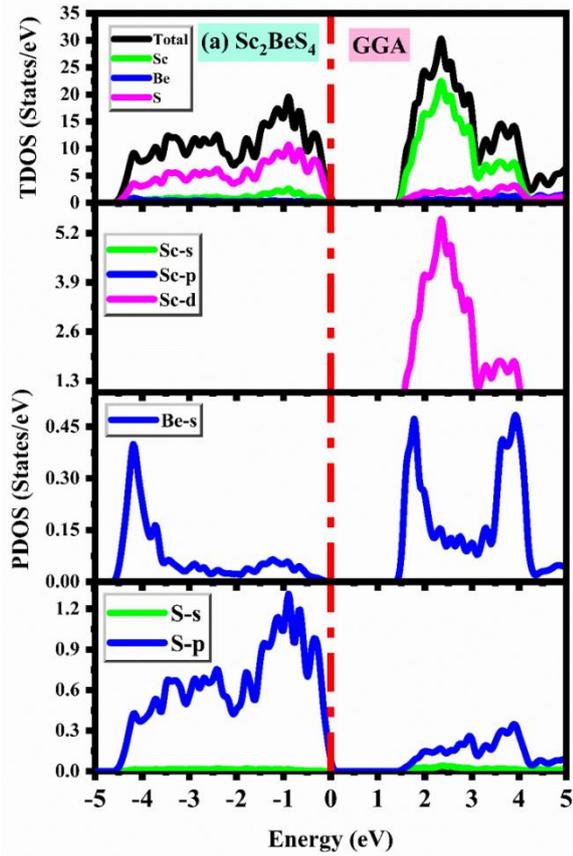
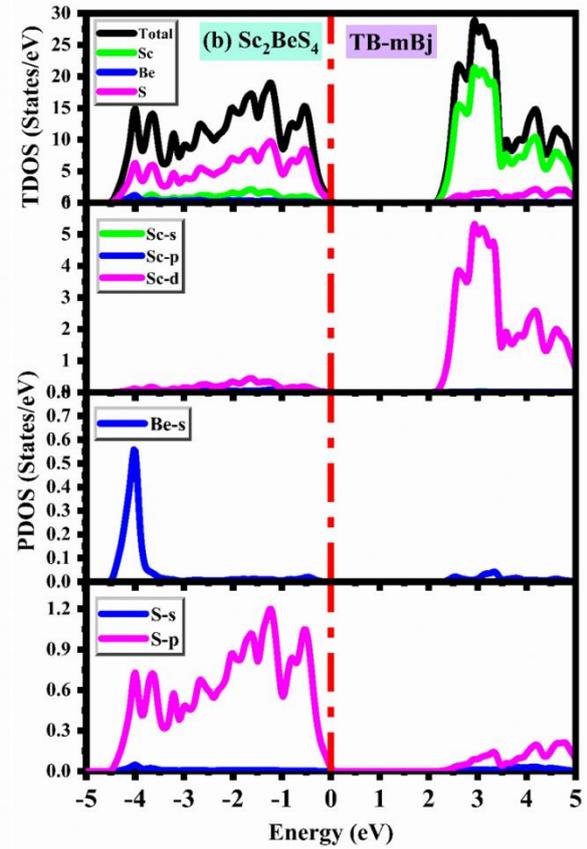

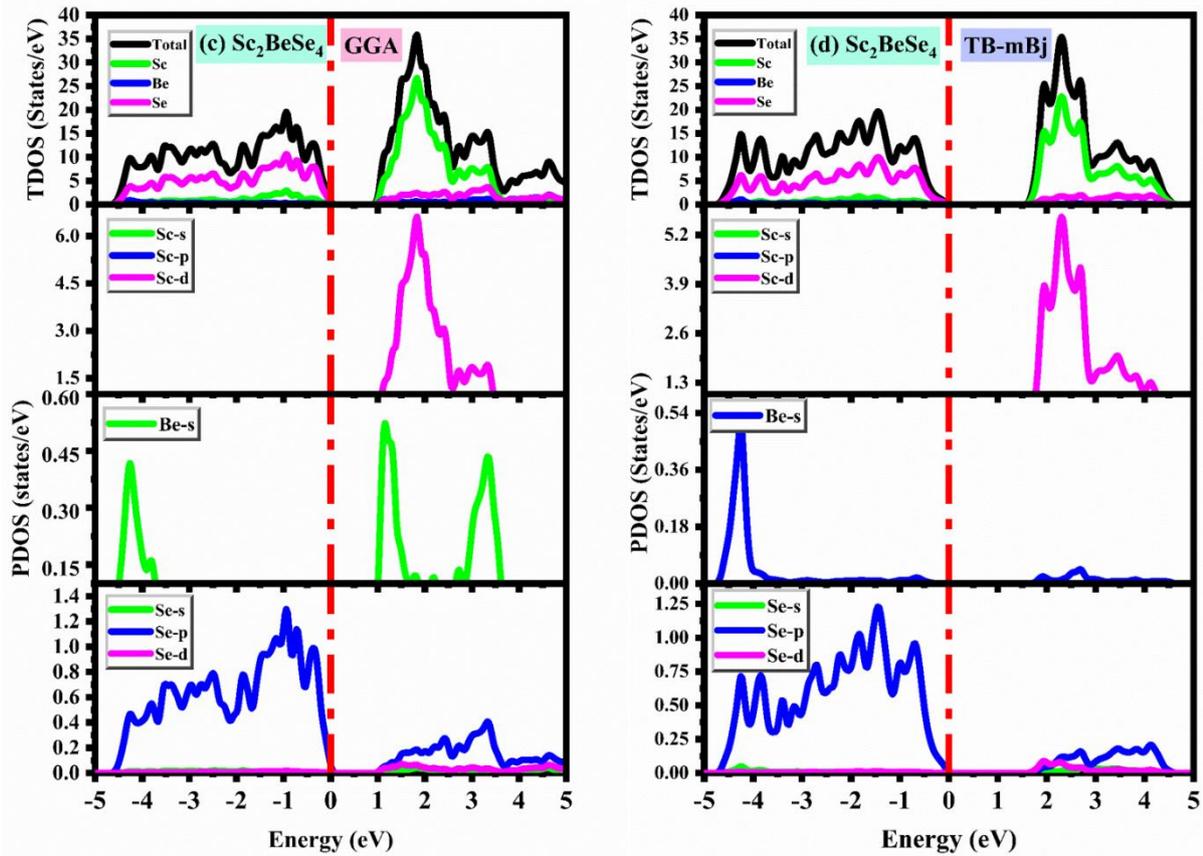

**Fig.4.** Illustration of the density of states (TDOS & PDOS) of (**a-b**) Sc$_2$BeS$_4$ and (**c-d**) Sc$_2$BeSe$_4$ in PBE-GGA and TB-mBJ approximations.

## 3.3. Optical Properties

The optical properties of a material are very important to understanding the light energy interaction with the materials. The study of these properties helps researchers to utilize the materials in solar cells and other photovoltaic applications. The major optical parameters of the materials Sc$_2$BeX$_4$ (X = S, Se) have been determined in the energy range of up to 14 eV, using the TB-mBJ approximations as the electronic bands structure computed by using TB-mBJ potential have closed band gaps to the experimentally measured, and depicted in Fig. 5. The optical properties of the materials give detailed information about their electronic nature and are dependent on their electronic band structures. The optical characteristics of the materials, such as dielectric functions, absorption, refractive index, energy loss function, and reflectivity, are examined for three regions, i.e., infra-red (IR), visible, and ultra-violet (UV). These optical spectra have different energy ranges. The energy range of infra-red lies in the range of 1.24 meV to 1.7 eV, the energy range of the visible region is 1.7 eV to 3.3 eV, and for the ultra-violet (UV), the energy range lies in between 3.3 eV to 124 eV [51, 52]. The dielectric function is an important complex function that explains the optical nature of the materials. Furthermore, the other optical parameters depend on the imaginary and real parts of the function can be understood from Eq. (6-9). The real part of the material explains the dispersion and polarization of the materials caused by the incident light. The real part $\varepsilon_1(\omega)$, for both materials, is displayed in Fig.5 (a). The static real portion $\varepsilon_1(0)$ of $Sc_2BeS_4$ and $Sc_2BeSe_4$ is 9 and 16.5, respectively. The study of static real parts of the materials follows Penn's Model [53], which states that the static real part of a material is proportional inversely to the energy band gap of the material. The energy band gap (TB-mBJ) of the material of $Sc_2BeS_4$ has a larger value (1.8 eV) and the static part has a smaller value (9), and of $Sc_2BeSe_4$ the energy band gap (TB-mBJ) has a smaller value, compared to $Sc_2BeS_4$, of (1.2 eV) and the static real part is (16.5). The biggest peak of $\varepsilon_1(\omega)$ for $Sc_2BeS_4$ and $Sc_2BeSe_4$ has approximate values at energies 2.4 eV and 1.7 eV, respectively. The real part plots for both materials suggest that the maximum peak is located in the visible region. The absorptive behavior of the materials can be understood by the analysis of the imaginary part $\varepsilon_2(\omega)$ of the materials. The observed static dielectric constants of 9 ($Sc_2BeS_4$) and 16.5 ($Sc_2BeSe_4$) fall within the expected range for materials with strong electronic polarizability. For comparison, lead halide perovskites like $MAPbI_3$ show $\varepsilon_1(0)$ values between 25-30 [54], while CZTS shows ~10–12 [55]. The relatively high values here are attributed to the significant hybridization between Sc-d and Se/S-p orbitals, enhancing polarizability.

The imaginary part $\varepsilon_2(\omega)$ of $Sc_2BeS_4$ and $Sc_2BeSe_4$ has been displayed in Fig. 5(b). The plots suggest that the threshold values of both materials are in complete accordance with the energy band gaps of the material. The maximum peaks of the $\varepsilon_2(\omega)$ of $Sc_2BeS_4$ and $Sc_2BeSe_4$ are observed to be located at 3 eV and 2.7 eV, respectively. The peaks of $\varepsilon_2(\omega)$ decrease gradually beyond the mentioned maximum energies of both materials. The plots clearly show that the maximum absorption of both materials lies in the visible region of the optical spectra. The absorption function $I(\omega)$ of the materials $Sc_2BeS_4$ and $Sc_2BeSe_4$ is displayed in Fig.5(c). The threshold values of both materials look similar to the imaginary part threshold values. The absorption $I(\omega)$ for both $Sc_2BeS_4$ and $Sc_2BeSe_4$ starts increasing from the visible region beyond the threshold value and reaches its maximum in the UV region of the optical spectra. The maximum absorption of both materials, $Sc_2BeS_4$ and $Sc_2BeSe_4$, takes place at an energy of 13.5 eV. Although the main solar spectrum lies below 5 eV, the presence of strong absorption in the deep UV range (13.5 eV) suggests possible applications in ultraviolet photodetectors and radiation shielding, broadening the application potential of $Sc_2BeX_4$ beyond photovoltaic uses.

The ability of a substance to disperse incident light and behave transparently is explained by its refractive index. The $n(\omega)$ represents the refractive index of the materials under study. The Fig. 5

(d) shows the refractive index n(ω) of both materials. The static values n (0) of the materials $Sc_2BeS_4$ and $Sc_2BeSe_4$ are 3.2 and 3.1, respectively. The n(ω) plots of both materials display that the peak values in both approximations are located in the visible region of the optical spectra. The energy loss function computed for the materials $Sc_2BeS_4$ and $Sc_2BeSe_4$ is represented by L(ω) and has been displayed in Fig. 5(e). In the infrared range, the L(ω) for both materials is negligible. It subsequently increases gradually with a rise in photon energy and reaches its maximum value for both materials located at 13 eV. The R(ω) reflectivity coefficient of the materials explains the surface roughness or smoothness of the materials. The computed R(ω) of $Sc_2BeS_4$ and $Sc_2BeSe_4$ has been displayed in Fig. 5(f). The static reflectivity R (0) of materials $Sc_2BeS_4$ and $Sc2BeSe4$ is 0.26 (26%) and 0.30 (30%), respectively. The reflectivity plots show several peaks for both materials. The first reflectivity peaks for both materials are observed to be in the visible region, and the maximum peaks are located in the UV region at an energy of 13.5 eV for both materials. The compounds exhibit metallic properties in the ultraviolet (UV) region and possess high reflectivity. These substances can be effectively used as a protective barrier against UV radiation and also as Bragg's reflectors [56]. The largest values of real parts, refractive index, imaginary parts, and absorption coefficient, and smaller values of the energy loss functions and reflectivity coefficients in the visible regions suggest that the materials are excellent candidates for solar cells and other photovoltaic applications.

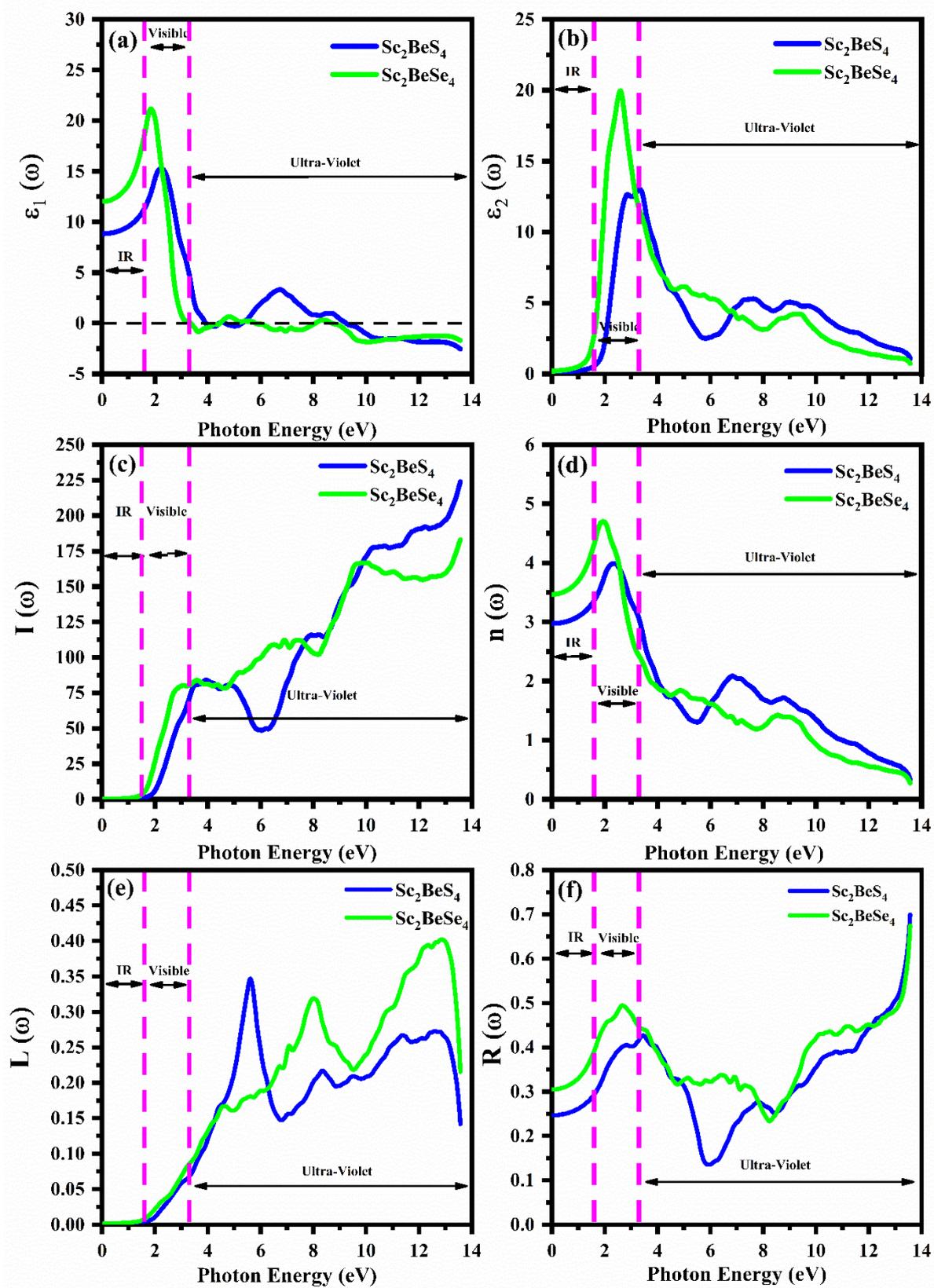

**Fig. 5.** Illustration of computed optical parameters of $Sc_2BeS_4$ and $Sc_2BeSe_4$ materials, including (**a**) the real part of the dielectric function; $\varepsilon_1(\omega)$ (**b**) the imaginary part of the dielectric function; $\varepsilon_2(\omega)$ (**c**) the absorption coefficient; $I(\omega)$ (**d**) the refractive index; $n(\omega)$ (**e**) the energy loss function; $L(\omega)$ (**f**) the reflectivity coefficient; $R(\omega)$.

## 3.4. Thermoelectric Properties

The thermoelectric transport behavior of $Sc_2BeS_4$ and $Sc_2BeSe_4$ was computed using the BoltzTraP code under the constant relaxation time approximation ($\tau = 5 \times 10^{-15}$ s) [57]. Although the electronic band structure primarily distinguishes whether a material is metallic, semiconducting, or insulating, it also serves as the foundation for predicting thermoelectric transport coefficients [58]. A key parameter in evaluating thermoelectric performance is the dimensionless figure of merit, ZT, which is derived from the electrical conductivity ($\sigma/\tau$), the Seebeck coefficient (S), and the electronic thermal conductivity ($\kappa_e/\tau$). The following sections present a detailed interpretation of the calculated properties in the 50–800 K range, as shown in Fig. 6(a–f).

Electrical conductivity is a measure of how easily charge carriers (electrons or holes) can move through a material under an applied electric field. High conductivity is desirable in thermoelectric generators and sensors where efficient charge transport is essential for output power [59]. The temperature-dependent electrical conductivity is plotted in Fig. 6(a). For both $Sc_2BeS_4$ and $Sc_2BeSe_4$, $\sigma/\tau$ rises steadily with temperature, a typical characteristic of semiconductors [59]. At low temperatures, the number of thermally excited carriers is small, resulting in lower conductivity. As the temperature increases, thermal excitation across the band gap generates additional carriers, causing a continuous rise in $\sigma/\tau$. At room temperature (300 K), $\sigma/\tau$ reaches $2.45 \times 10^{18}$ $(\Omega\,m\,s)^{-1}$ for $Sc_2BeS_4$ and $1.91 \times 10^{18}$ $(\Omega\,m\,s)^{-1}$ for $Sc_2BeSe_4$. The slightly higher conductivity of the S-based compound indicates enhanced carrier mobility or a more favorable band-edge configuration. The linear increase of $\sigma/\tau$ up to 800 K reflects dominant phonon scattering, which is commonly observed in crystalline p-type semiconductors.

The Seebeck coefficient represents the thermoelectric voltage generated in response to a temperature gradient, indicating the strength and polarity of carrier transport. Materials with high S values are widely applied in thermoelectric sensors, waste heat recovery devices, and p–n junction-based power generators [60]. The Seebeck coefficient, displayed in Fig. 6(b), quantifies the thermoelectric power and the ability of a material to convert a temperature gradient into an electrical voltage [60]. Both compounds exhibit **positive Seebeck coefficients**, confirming their p-type behavior. From 100 K to 400 K, S increases from approximately $1.3 \times 10^{-4}$ V/K to about $2.5 \times 10^{-4}$ V/K, reflecting enhanced hole diffusion as more carriers are thermally activated. Beyond 400 K, S remains nearly constant, indicating that the Fermi level is stabilized relative to the band edges and that further thermal excitation does not significantly alter the thermopower. The p-type nature inferred from S aligns well with the band structure analysis and suggests that these materials can maintain stable thermoelectric voltage generation at high temperatures [33].

Molar specific heat capacity defines the amount of heat energy required to raise the temperature of one mole of a material by one degree. High and stable $C_v$ values are vital for thermal management in industrial ceramics, thermal barrier coatings, and materials operating in high-temperature environments [61]. The molar specific heat capacity, shown in Fig. 6(c), represents the heat storage ability of the lattice. Both compounds exhibit a linear increase in $C_v$ with temperature up to around 350 K, after which the slope gradually decreases as the Dulong–Petit limit is approached. This trend signifies that most phonon modes become thermally populated at elevated temperatures, leading to saturation of the heat capacity. The almost identical values of $C_v$ for $Sc_2BeS_4$ and $Sc_2BeSe_4$ imply that the substitution of S by Se does not drastically modify the vibrational entropy or lattice dynamics. The linear regime at low temperatures closely follows Debye's law, affirming the crystalline quality and strong lattice interactions of these materials [61]. The high $C_v$ values at elevated temperatures further support their thermal stability, which is crucial for thermoelectric operation.

Electronic thermal conductivity quantifies the ability of free carriers to transport heat through a material. Controlled $\kappa_e$ is important in thermoelectric devices, where excessive thermal conductivity reduces efficiency, and in electronics cooling, where effective heat dissipation is needed [62]. The calculated electronic thermal conductivity, presented in Fig. 6(d), increases steadily with temperature beyond approximately 150 K. This rise is attributed to enhanced phonon–electron interactions and the growing contribution of thermally excited carriers to heat transport [62]. At 800 K, $\kappa_e/\tau$ attains a value of $37 \times 10^{14}$ W/m K s for $Sc_2BeS_4$ and $29.6 \times 10^{14}$ W/m K s for $Sc_2BeSe_4$. The slightly higher $\kappa_e/\tau$ of the sulfur compound suggests more efficient charge and heat transport channels. The temperature dependence indicates that as thermal energy increases, lattice vibrations intensify and free carriers gain mobility, which collectively enhance the thermal conductivity.

Power factor, defined as S²σ, is a combined parameter indicating how efficiently a material converts heat into electrical power. A high PF is crucial for high-performance thermoelectric generators, Peltier cooling devices, and energy harvesters [63]. The power factor, defined as S²σ and depicted in Fig. 6(e), is a direct measure of the thermoelectric conversion efficiency [63]. Both materials show a near-linear increase in PF with temperature, signifying improved electrical performance at elevated operating conditions. At 300 K, the PF of $Sc_2BeS_4$ reaches approximately $1.25 \times 10^{11}$ W/K² m s, while $Sc_2BeSe_4$ exhibits a slightly lower value of about $1.10 \times 10^{11}$ W/K² m s. The steady enhancement of PF up to 800 K is consistent with the simultaneous increase of σ/τ and the relatively high Seebeck coefficient over this temperature span.

The dimensionless figure of merit, ZT, is the key metric used to evaluate the overall efficiency of a thermoelectric material, combining S, σ, and κ into one parameter. Materials with high ZT values are extensively employed in solid-state cooling, automotive waste heat recovery, and power generation in aerospace and remote sensing devices [64]. The overall thermoelectric efficiency is described by the figure of merit (ZT), shown in Fig. 6(f), which combines electrical conductivity, Seebeck coefficient, and thermal conductivity into a single dimensionless parameter [64]. At 300 K, ZT is approximately 0.70 for $Sc_2BeS_4$ and 0.78 for $Sc_2BeSe_4$, while at 800 K the values rise to about 0.80 and 0.79, respectively. The consistent improvement of ZT with temperature reflects the enhanced electrical transport properties and favorable balance with thermal conductivity. The high ZT values, particularly at elevated temperatures, underscore the potential of both $Sc_2BeS_4$ and $Sc_2BeSe_4$ for high-performance thermoelectric applications, where the conversion of waste heat into electricity is desirable.

The overall comprehensive temperature-dependent analysis of σ/τ, S, $C_v$, $κ_e$/τ, PF, and ZT confirms that $Sc_2BeS_4$ and $Sc_2BeSe_4$ are promising p-type thermoelectric materials with excellent stability, strong lattice integrity, and high thermoelectric efficiency across a wide temperature range. The sulfur-based compound demonstrates slightly higher conductivity and thermal transport, whereas the selenium-based counterpart exhibits comparable Seebeck performance and slightly higher ZT at lower temperatures, making both materials viable candidates for next-generation thermoelectric devices.

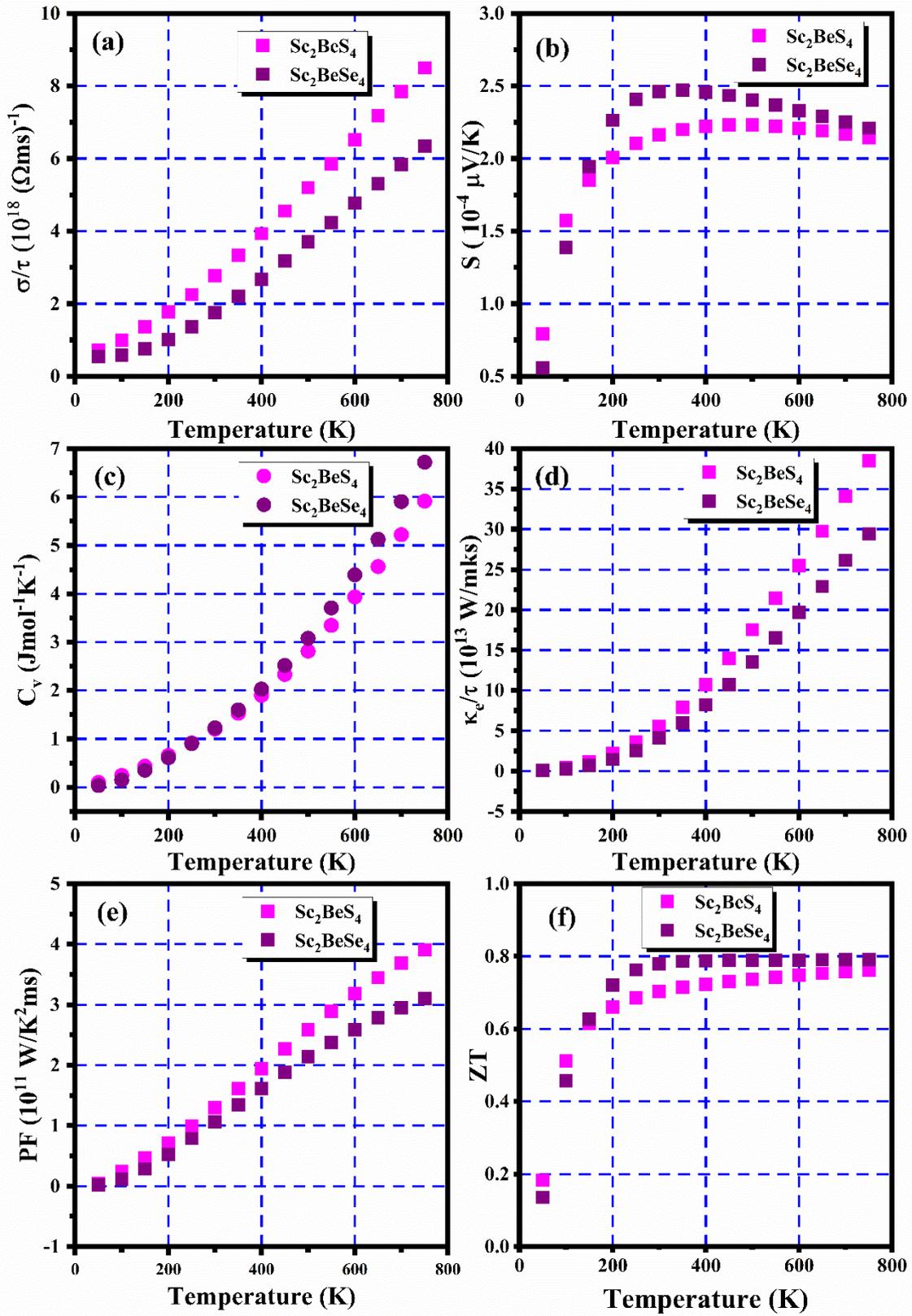

**Fig.6.** Illustration of thermoelectric properties of $Sc_2BeS_4$ and $Sc_2BeSe_4$ materials, include (**a**)Electrical Conductivity ($\sigma/\tau$), (**b**)Seebeck Coefficient (S), (**c**) Specific heat capacity (Cv), (**d**) Electronic Thermal Conductivity ($\kappa_e$), (**e**) Power Factor (PF) and (**f**) ZT.

### 3.5. Thermodynamic Properties

Thermodynamic properties analysis of materials is a fundamental science that helps process design and provides theoretical foundations for many other scientific disciplines. To explore the thermodynamic properties of the $Sc_2BeX_4$ (X = S, Se) materials quasi-harmonic Debye model [65] in CASTEP computational code, is used to compute the thermodynamic characteristics like the Enthalpy, Free Energy, Entropy, and specific heat capacity at constant volume $C_v$, displayed in Fig. 7, also Debye Temperature is computed and displayedin Fig. 8, over the 0 to 1000 K temperature range. The computed enthalpy in eV of both materials has been plotted in Fig. 7 (a). Enthalpy is the sum of U, the internal energy of the materials, and the product of P, pressure, and V, Volume. The enthalpy is the measure of energy required to create a material [66]. The plots suggest a linear increase in the enthalpy with the rise in temperature beyond 50 K, which predicts the rise of internal energy and PV with temperature rise. According to thermodynamics, a system's free energy is the amount of energy that can be used to accomplish work while the temperature and pressure are both constant. The ability to forecast the path of chemical reactions and phase shifts makes Gibbs free energy especially significant. The Gibbs free energy of the materials has been plotted in Fig. 7 (b) and shows a linear decline for both materials. A linear decrease in free energy generally suggests that the entropic contribution of the material dominates while maintaining a level of stability as temperature increases. This phenomenon frequently occurs in systems where vibrational entropy predominantly influences entropy at elevated temperatures, in the absence of phase transitions or notable structural alterations. A linear trend typically suggests that there are no phase transitions taking place within the temperature range of 0 K to 1000 K. Phase transitions frequently result in sharp changes in free energy, driven by sudden shifts in entropy or enthalpy. Entropy is crucial for comprehending the characteristics and behavior of materials. Entropy is an essential indicator for assessing how matter and energy change in different materials and processes [67]. The entropy of the materials has been displayed in Fig. 7 (c). The plots show a linear increase in the entropy, which suggests a stable material with constant heat capacity, and no phase transitions up to 1000 K. The specific heat capacity at constant volume, $C_V$, increases up to 500 K and remains constant beyond the mentioned temperature. The constant behavior of the $C_V$ beyond 500 K suggests the thermodynamic stability of the materials. Overall, thermodynamic analysis suggests the thermodynamic and phase stability of the materials. The Debye temperatures for $Sc_2BeS_4$ and $Sc_2BeSe_4$ were found to be approximately 420 K and 360 K, respectively. These moderate values reflect relatively soft lattice dynamics and suggest low lattice thermal conductivity, a desirable property for thermoelectric applications where phonon scattering is beneficial.

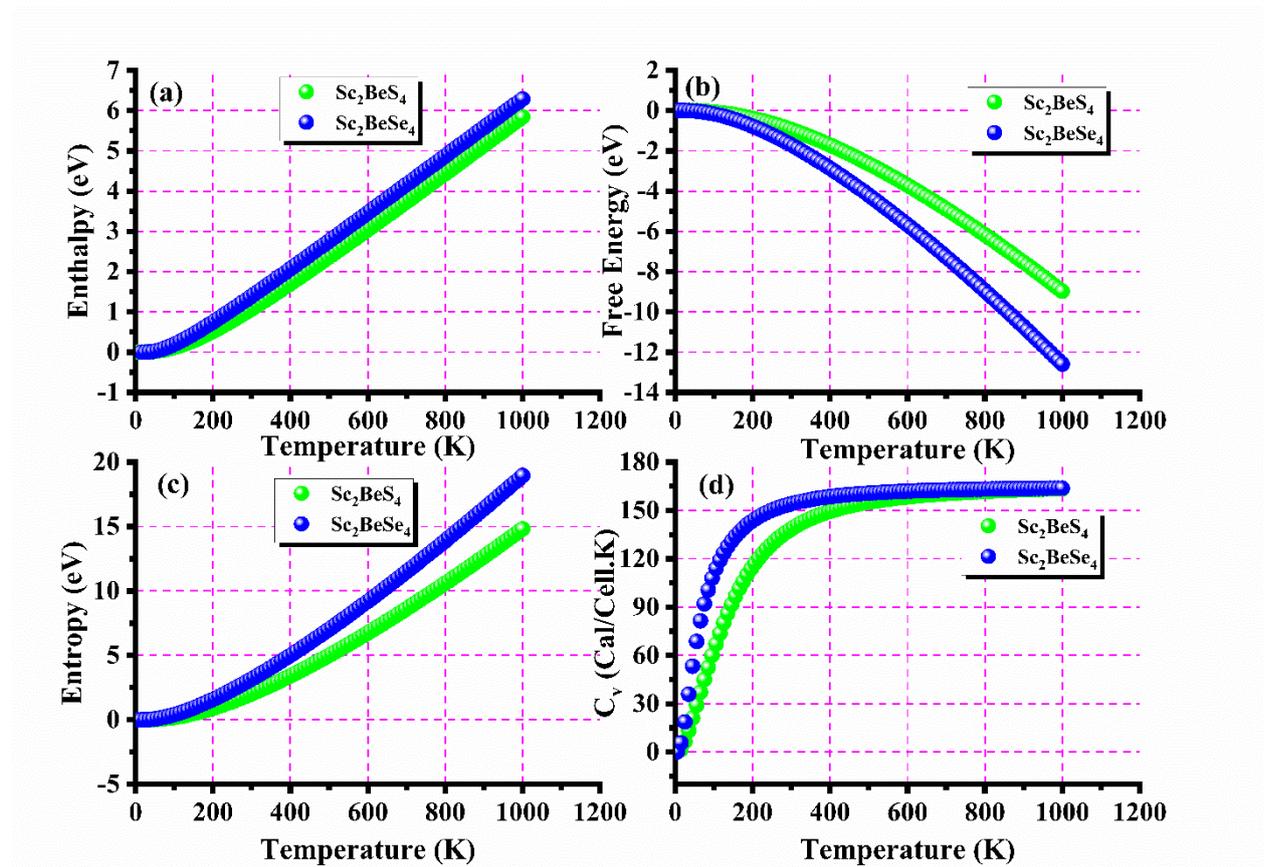

**Fig. 7:** Thermodynamic Characteristics of the Materials (**a**) Enthalpy (**b**) Gibbs's Free Energy (**c**) Entropy (**d**) Specific heat capacity at constant volume Cv.

## 4. Conclusion

The DFT study of $Sc_2BeX_4$ (X = S, Se) ternary chalcogenide materials in a tetragonal phase with space group number 122 (I-42D) has been conducted to predict their thermoelectric and optoelectronic nature, using TB-mBJ and PBE-GGA approximations. The electronic nature of the materials has been investigated by computing their electronic band structures and density of states in both approximations. The electronic analysis suggests the direct band gap semiconducting nature of the materials. The energy band gaps of the material $Sc_2BeS_4$ are 1.4 eV (PBE-GGA) and 1.8 eV (TB-mBJ), and of $Sc_2BeSe_4$ are 0.8 eV (PBE-GGA) and 1.2 eV (TB-mBJ). The DOS suggests that the major contribution to the electrical conductivity of $Sc_2BeS_4$ and $Sc_2BeSe_4$ is due to S-p and Se-p states, respectively. The energy band gap of the materials suggests that the materials can be used in solar cells for harvesting solar energy into useful electrical energy. The utilization of the materials in solar cell applications has been analyzed by computing the optical characteristics of the materials. The photonic characteristics, like parts of the dielectric function, the reflectivity coefficient, the absorption, refractive index, energy loss function, and reflectivity have been computed in TB-mBJ approximations in the energy range of 0 to 14 eV. The materials appear ideal candidates for solar cells and other photovoltaic applications, as shown by the maximum values of real parts, refractive index, imaginary parts, and absorption coefficient, and smaller values of energy loss functions and reflectivity coefficients in the visible areas of the optical spectra. The thermoelectric parameters have been computed to investigate the thermoelectric nature of the materials. The specific heat capacity, Seebeck coefficient, and thermal and electrical conductivity have been computed using the BoltzTrap code. The Seebeck coefficient suggests the p-type semiconducting nature of both materials. The specific heat capacity shows that the materials are thermodynamically stable. The thermal and electrical conductivity increase linearly and have the highest value at 800 K. The power factor (PF) and ZT have also been computed to understand the thermoelectric performance and efficiency of the materials to convert the heat energy into electrical energy. The results explain that the PF of the materials at 300 K is 1.25 x $10^{11}$ W/K$^2$ms and 1.1 x $10^{11}$ W/K$^2$ms for $Sc_2BeS_4$ and $Sc_2BeSe_4$, respectively. The ZT values of $Sc_2BeS_4$ are 0.7 (70%) at 300 K and 0.8 (80%) at 800 K, and of $Sc_2BeSe_4$ are 0.78 (78%) at 300 K and 0.79 (79%) at 800 K. The results of these parameters demonstrate that the materials are great choices for applications using thermoelectric devices. Thermodynamic properties of the materials include enthalpy, Gibbs' free energy, entropy, and specific heat capacity at constant volume, Cv. The results show a linear increase with temperature in enthalpy, specific heat capacity at constant volume, Cv, and entropy, and a linear decrease in Gibbs Free energy. The thermodynamic analysis of the materials supports the thermodynamic and phase stability of the materials.

**Conflict of Interests**

The authors declare that there are no conflicts of interest regarding the publication of this paper. All authors have contributed to this work according to the academic and research standards, and there are no competing interests, financial or otherwise, that could have influenced the outcomes of this study.